\documentclass[9pt]{extarticle}
\usepackage{cite}
\usepackage{amsmath}
\usepackage{graphicx}
\usepackage{subfigure}
\usepackage{verbatim}
\usepackage{widetable}
\usepackage[justification=justified,singlelinecheck=false]{caption}
\usepackage{textgreek}
\usepackage{spconf}
\usepackage{makecell}
\usepackage{multirow}
\usepackage{url}
\renewcommand{\vec}[1]{\mathbf{#1}}
\graphicspath{{figures/}}

\DeclareGraphicsExtensions{.pdf}

\title{FPGA BASED IMPLEMENTATION OF DEEP NEURAL NETWORKS USING ON-CHIP MEMORY ONLY}
\name{Jinhwan Park and Wonyong Sung\thanks{This work was supported in part by the Brain Korea 21 Plus Project and the National Research Foundation of Korea (NRF) grant funded by the Korea government (MSIP) (No. 2015R1A2A1A10056051).}}
\address{
Department of Electrical  and Computer Engineering\\
Seoul National University\\
Seoul 151-744 Korea\\
Email: jhpark@dsp.snu.ac.kr, wysung@snu.ac.kr}

\begin{document}
\maketitle

\begin{abstract}

Deep neural networks (DNNs) demand a very large amount of computation and weight storage, and thus efficient implementation using special purpose hardware is highly desired.  In this work, we have developed an FPGA based fixed-point DNN system using only on-chip memory not to access external DRAM. The execution time and energy consumption of the developed system is compared with a GPU based implementation.  Since the capacity of memory in FPGA is limited, only 3-bit weights are used for this implementation, and training based fixed-point weight optimization is employed. The implementation using Xilinx XC7Z045 is tested for the  MNIST handwritten digit recognition benchmark and a phoneme recognition task on TIMIT corpus.  The obtained speed is about one quarter of a GPU based implementation and much better than that of a PC based one.  The power consumption is  less than 5 Watt at the full speed operation resulting in much higher efficiency compared to GPU based systems. 

\end{abstract}



%


\section{Introduction}
\label{sec:intro}

Feed-forward deep neural networks (DNNs) show quite good performance in speech and pattern recognition applications \cite{hinton2006fast,hinton2006reducing}.  Real-time implementation of feed-forward deep neural networks demand a very large number of arithmetic and memory access operations, thus DNNs are usually implemented using GPUs (Graphics Processing Units) \cite{chen2014fast,zhang2013optimized}. GPU based implementations consume large power exceeding 100 Watt\cite{nvidiapower}. In addition, a GPU based system needs a PC that occupies a large space, which may not be suitable for embedded applications requiring small foot-print units.

There are several previous works on VLSI and FPGA based implementation of a DNN or a CNN (Convolutional Neural Network).  The system in \cite{farabet2010hardware} stores the weights at the external DRAM, and can configure the algorithm fairly flexibly.  However, this system demands a large number of external memory accesses, and the throughput is fairly limited.  A full custom VLSI that employs thousands of processing units and stores the weights at the on-chip memory was developed in \cite{kim2014x1000}.  This custom VLSI based system can achieve a very high throughput and consumes small power, but is not flexible.

In a general feedforward deep neural network with multiple hidden layers, each layer $k$ has a signal vector $\vec{y}_k$, which is propagated to the next layer by multiplying the weight matrix $\vec{W}_{k+1}$, adding biases $\vec{b}_{k+1}$, and applying the activation function $\phi_{k+1}(\cdot)$  as follows:
  \begin{align}
    \label{eq:1}
      \vec{y}_{k+1}=\phi_{k+1}(\vec{W}_{k+1}\vec{y}_k + \vec{b}_{k+1}).
\end{align}     
A deep neural network usually employs one to six hidden layers. One of the most general activation functions is the logistic sigmoid function defined as 
  \begin{align}
    \label{eq:2}
    \sigma(x)=\frac{1}{1+e^{-x}}.
\end{align} 
In fully-connected feedforward deep neural networks, each weight matrix between two layers demands parameters whose capacity is determined by the product of units in the anterior  and the posterior layers. Considering a DNN employing hidden layers with 1,000 units, each weight matrix demands about one million parameters. This means that a few million weights are needed for the implementation of a typical DNN and more than 10MB of memory is needed when the weights are represented in the floating-point format. The number of output signals and that of biases are both  proportional to the layer size. The signal word-length affects the complexity of arithmetic units and interconnection networks.

Reducing the complexity of neural networks using quantization or pruning has been studied much\cite{misra2010artificial,fiesler1990weight,
moerland1997neural,tang1993multilayer,han2015learning,yu2012exploiting}. Instead of direct weight quantization, retraining with backpropagation was developed in \cite{kim2014x1000,hwang2014fixed,anwar2015fixed}. The designed networks employed usually 2-8 bits for the weights and represented the signals in analog or high precision fixed-points using more than 7 bits. Recently, a research work that tries to increase the sparseness of the weights by pruning out small valued ones has been developed in order to reduce the model size and the execution time with a CPU \cite{han2015learning,yu2012exploiting}. 

In this paper, we have developed a DNN using an off-the-shelf Xilinx FPGA, XC7Z045 aiming for design flexibility, high throughput with thousands of processing units, and low-power consumption with virtually no DRAM accesses by storing all the weights on on-chip memory. Since the on-chip memory of an FPGA is the most limited resource for the implementation of a DNN, we employ the training based weight quantization scheme and achieve a quite good performance only with 3 bits for the weight representation \cite{yu2012exploiting}. A handwritten digit recognition and phoneme recognition problems are implemented in this FPGA.  It is, however, takes less than a few hours to develop a DNN with a different network configuration. 

This paper is organized as follows. In Section 2, we describe the algorithm and the architecture for this FPGA based DNN system. Section 3 describes the implementation detail of this work. The experimental results are shown in Section 4. Concluding remarks are shown in Section 5. 

\section{Algorithm And Architecture Optimization}
\label{sec:dnn_conv}

\subsection{Fixed-Point DNN design}
\label{subsec:sr_kws}

The DNN for handwritten digit recognition has three hidden layers, and the layer configuration is 784-1022-1022-1022-10. The input of the DNN is a $28\times 28$ (=784) pixel image represented in 8-bit gray scale.  The output corresponds to the likelihood of each digit. This algorithm needs approximately 3 million weights, which is the sum of $784\times1022$ + $1022\times1022$ + $1022\times1022$ + $1022\times10$.

The DNN for phoneme recognition has four hidden layers. Each of the hidden layers has 1022 units. The input layer of the network has 429 units to accept 11 frames of MFCC (Mel-frequency cepstral coefficient) parameters. The output is the likelihood of 61 phonemes.

The FPGA used for this implementation is Z-7045 of Xilinx.  This FPGA contains 218K LUTs (Look Up Tables), 2.18 MB of block RAM (BRAM), and 900 digital signal processing (DSP) blocks in addition to dual A9 ARM Cortex CPUs.  Even if we employ 8 bits for representing each weight, the BRAM cannot accommodate all the weights and thus many DRAM accesses are needed.

%

In order to store all the weights on the FPGA, we apply fixed-point optimization for the weights, and successfully reduce the word-length into 3 bits. The weights are trained at the off-line and downloaded to this system. The network training consists of three steps; the first one is an ordinary floating-point training, the second is the optimal uniform quantization minimizing the error in L2 norm, and the final step is the retraining with fixed-point weights. 

For digit classification, the network is pre-trained with unsupervised greedy RBM (Restricted Boltzmann Machine) learning. Each layer is pre-trained by 50 epochs of 1-step contrastive-divergence based stochastic gradient descent with the mini-batch size of 100, the learning rate of 0.1, and the momentum of 0.9. Then, we ran 100 epochs of the backpropagation with stochastic gradient descent using the mini-batch size of 100, the fixed learning rate of 0.1, and the momentum of 0.9. MNIST dataset is used as the training set.

For phoneme recognition, each layer is pre-trained by 50 epochs of stochastic gradient descent with the mini-batch size of 128, the learning rate of 0.05, and the momentum of 0.9. We ran 100 epochs of the backpropagation using the mini-batch size of 128, the fixed learning rate of 0.05, and the momentum of 0.9 for fine tuning. TIMIT database is used for training. Dropout is applied for training both of networks. 

After obtaining the floating-point weights, we apply fixed-point optimization that employs retraining after quantization \cite{hwang2014fixed}. The same training parameters are used for retraining weights. The weights of fixed-point DNN employ 3 bits for the input and hidden layers, and 8 bits for the output layer that is more sensitive to quantization. By this fixed-point optimization, we can obtain the miss-classification rate (MCR) of 1.08\%\cite{hwang2014fixed}. Note that the DNN with 1022 floating-point units show the MCR of 1.06\%, and the DNN with 900 floating-point units yields the MCR of 1.12\%. For the phoneme recognition, the phoneme error rate of the floating-point DNN is 27.81\%, while that of the 3-bit fixed-point DNN is 28.39\%. 


\subsection{Parallel-serial architecture}
\label{subsec:kwf_hmm}

The DNN for digit recognition needs approximately 3K neuron-like processing units: 1022 units for implementing hidden layer1, 1022 for hidden layer2, 1022 for hidden layer3, and 10 for the output layer.  Each neuron-like processing unit conducts approximately 1022 multiply and add operations to process one input image.  As a result, the total amount of computation for processing one image is approximately 3 million operations. 

\begin{figure}[!t]
\centerline
{
	\includegraphics[angle=0, width=3.2in]{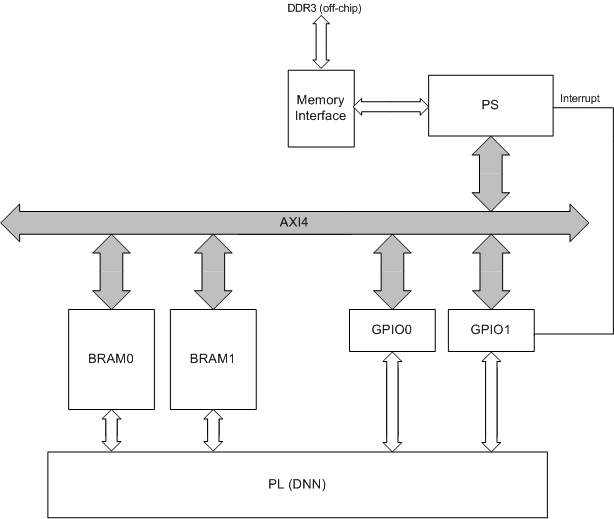}
\
}
\caption{System Overview.}
\label{fig_sr}
\end{figure}

In order to trade the throughput and the hardware resources, we employ the parallel-serial architecture.  In this design, each layer contains 1022 processing units (PUs), which means parallel PU operations, and each PU computes the output in 1022 clock cycles, which is the serial processing in each PU.  At each clock cycle, the PU conducts only one multiply-add operation.  In fact, since the weights are quantized to 3 bits in the input and hidden layers, the PUs do not need multipliers. However, the synthesis shows that this design results in overuse of LUTs.  As a result, the number of PUs is halved and the processing time becomes doubled as 2044 + some overhead cycles.  With the system clock frequency of 100 MHz, this design can compute approximately 48.9 K outputs per second.  

\section{FPGA BASED IMPLEMENTATION}
\label{sec:int_kws}

Figure 1 shows the overall block diagram of the system.  The processing system (PS) and the programmable logic (PL) communicate through the general purpose IO (GPIO).  For digit recognition, one hundred images are recognized at each operation in this design.  However, the number of images to process at each batch can be changed easily.  The operations are performed as follows.

\begin{itemize}
  \item PS moves 100 handwritten images stored at the DRAM to BRAM0 (or BRAM1 alternatively), and activates the start signal in GPIO0. 
  \item PL conducts the recognition of 100 images, writes the result to BRAM0 (or BRAM1 alternatively), and then activates the `Done' signal in GPIO0 when the recognition process is finished. This process takes approximately 200K cycles.
\end{itemize}

As the operation is explained in the above, the PS and PL work in an interleaved manner using BRAM0 and BRAM1. The system for phoneme recognition operates in a similar way. 
\begin{figure}[!t]
\centerline
{
	\includegraphics[angle=0, width=3.2in]{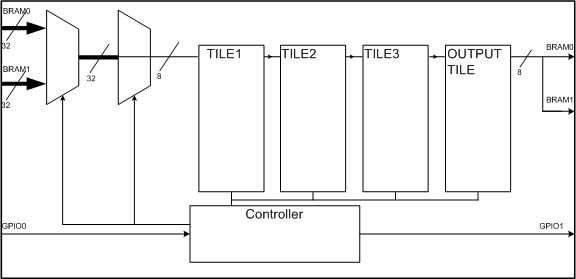}
}
\caption{Architecture of deep neural network circuit.}
\label{fig_gmm_vs_dnn}
\end{figure}
\begin{figure}[!t]
\centerline
{
	\includegraphics[angle=0, width=3.2in]{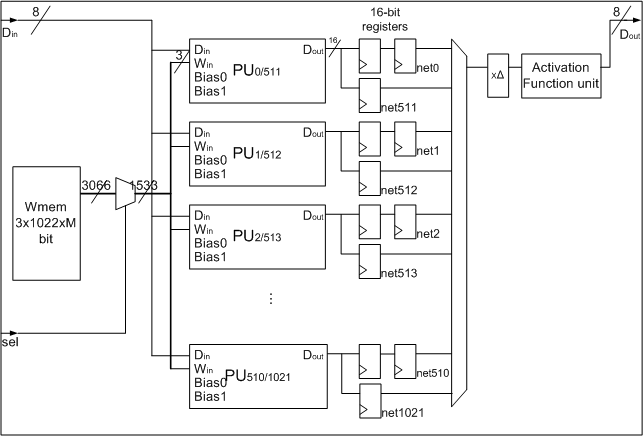}
}
\caption{Structure of a tile. M = 1022 for this figure.}
\label{fig_int_kws}
\end{figure}

Figure 2 shows the overall architecture of the circuit, where each tile implements a layer of the DNN algorithm.  The input and output of the tile employ 8-bit word-length for signal representation.
The structure of a tile is shown in Fig. 3. The tiles used for hidden layers have 511 (=1022/2) PUs each.  Since each PU conducts the operation for two nodes, it has two output registers.  Each tile contains BRAM for weight storage.  Only one activation function is implemented in each tile.

The output tile contains a small number of output nodes, and employs 8-bit weight values. Since the number of nodes at the output tile is small, each PU conducts
the operations that correspond to the function of one neuron. The result of the
output node is compared to find the recognized digit or phoneme.

\begin{figure}[!t]
\centerline
{
	\includegraphics[angle=0, width=3.2in]{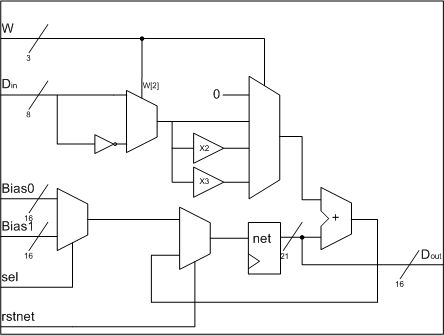}
}
\caption{Structure of a processing unit.}
\label{fig_int_kws}
\end{figure}

\begin{figure}[!t]
\centerline
{
	\includegraphics[angle=0, width=3.2in]{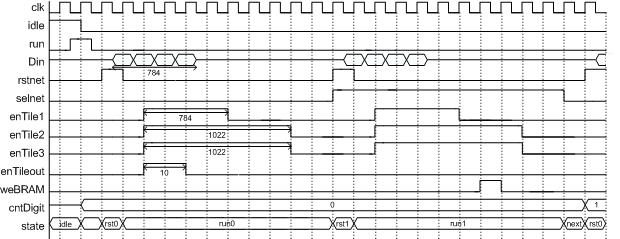}
}
\caption{Timing diagram of control signals.}
\label{fig_int_kws}
\end{figure}
Figure 4 shows the structure of each PU. Each PU multiplies the input with the weight of -3 to 3, and adds the multiplied result to the partial sum stored in the accumulation register.  According to the value of the weight, -3Din, -2Din, -Din, 0, Din, 2Din, or 3Din is added to the partial sum.  This logic is implemented using LUTs without consuming any DSP block.  The word-length of the adder in the PU is 16 bits.  The final output of each PU is multiplied with the coefficient $\Delta$, then the result (8-bit) is delivered to the next layer through the activation function. 

The activation function is also implemented using combinational logic circuits. The minimized sum-of-product representation is used for the sigmoid function as proposed in \cite{tommiska2003efficient}. We employ 8 bits for both input and output signals of the activation function.

The timing diagram of the operation is shown in Fig.  5.  At first, `rstnet' signal initializes the PUs.  Since the value of `selnet' is 0, each PU is initialized as its Bias0 value. After then, each tile receives 511 inputs from the previous tile and conducts the weight and accumulation operations using all the PUs in the tile.  Note that it takes 2 clock cycles for processing one input because one PU implements two nodes.  After then, the `selnet' becomes 1, and the next 511 input is processed using Bias1 and the corresponding weights.  After finishing this, the `cntDigit' increases, and the next image is processed at the same way.  When the cntDigit becomes 100, `fin' signal is generated and the PL becomes in the idle state.  In order to recognize one image, this circuit consumes 2063 (=2$\times$1022 + overhead) clock cycles.   

\section{Experimental Results}
\label{sec:exp_result}

This work is designed using Xilinx Vivado (v.2014.4), and implemented on a Xilinx ZC706 evaluation board.  This board contains XC7Z045 FPGA that contains both ARM CPUs and configurable logic circuits.  The CPU frequency is 800 MHz. The Programmable Logic (PL) operates at 172 MHz for digit classification and 140 MHz for phoneme recognition. The FPGA resource utilization is shown in Table 1, where the designs with and without DSP slices are compared. Using DSP slices does not reduce the demand of FFs (flip-flops) and LUTs much. For comparison, resource utilization of the implementation with 8-bit fixed-point weights is given. It consumes almost all of LUTs and entire DSP slices on the FPGA chip because 8-bit weights need hardware multipliers for the implementation. The 8-bit representation for weights cannot be implemented on this FPGA because the BRAM capacity is not sufficient. 

\begin{table}[!t]
\normalsize \centering
\renewcommand{\arraystretch}{1.6}
\caption{FPGA resource utilization for digit recognition with different weight precision.}
\label{tbl:dnn_kws_nkws}
\begin{tabular}{c | c | c | c | c}
\noalign {\vspace{1em}}
\noalign{\hrule height 2pt }
{Resource}		&	{FF }	&	{LUT } & {BRAM} & {DSP} \\
\hline
{3-bit without DSP}		&	{130237}	&	{124862} 	&	{323}		  & {0}\\
{3-bit with DSP}	&	{130802}		&	{121173} 		&	{323} 		 & {900}\\
{8-bit fixed point}	&	{136677}	&	{213593}	&	{750.5}          & {900}\\
\hline
\hline
{Available}	&	{437200}	&	{218600}	&	{545}	&	{900}	\\
\noalign{\hrule height 2pt }
\end{tabular}
\end{table}

\begin{table}[!t]
\normalsize \centering
\renewcommand{\arraystretch}{1.6}
\caption{FPGA resource utilization for phoneme recognition.}
\label{tbl:dnn_kws_nkws}
\begin{tabular}{c | c | c | c | c}
\noalign {\vspace{1em}}
\noalign{\hrule height 2pt }
{Resource}		&	{FF }	&	{LUT } & {BRAM} & {DSP} \\
\hline
{Used}		&	{161923}	&	{137300} 	&	{378}		  & {0}\\
\hline
\hline
{Available}	&	{437200}	&	{218600}	&	{545}	&	{900}	\\
\noalign{\hrule height 2pt }
\end{tabular}
\end{table}

The time spent for recognizing 10,000 images is measured as 142~msec. This implementation can process about 70,000 images at each second. If weights were stored in DRAM, the  memory bandwidth of 630 Gbit/sec (=3$\times$3M$\times$70000) is required to achieve this throughput. The memory bandwidth of the DRAM in Xilinx ZC706 board is 102.4 Gbit/sec. It is possible to reduce the number of DRAM accesses by processing multiple images at a time, which however demands extra internal registers for storing intermediate results and incurs more delay.

For the phoneme recognition application, the time spent for recognizing 10,000 frames is measured as 151~msec. This can process about 66,000 frames per second. Since a real-time speech recognition mostly uses 10 ms as for the frame size, this implementation result translates 660 times of speed-up over the real-time.  One FPGA can conducts phoneme recognition needed for real-time speech recognition of over 600 people.  

The power consumption shown at the simulation tool is shown in Table 3.   The real power measurement of the evaluation board is 11.4 Watt and 13.1 Watt for digit and phoneme recognition, respectively, which includes all the power consumption in the peripherals and the power circuit.  As shown in this table, the power consumed in the recognition circuit is about 5 Watt for obtaining the throughput of 70,000 images/sec. This translates 71~\textmu J for one image recognition.  

\begin{table}[!t]
\normalsize \centering
\renewcommand{\arraystretch}{1.6}
\caption{On-Chip power consumption (Watt) estimated by Xilinx Vivado.}
\label{tbl:dnn_kws_nkws}
\begin{widetable}
{\columnwidth}{c | c | c }
\noalign {\vspace{1em}}
\noalign{\hrule height 2pt }
{Usage}		&	{Digit classification }	&	{Phoneme recognition }  \\
\hline
{Clocks}	&	{0.579}	&	{0.528} 	\\
{Signals}	&	{1.202}	&	{3.559} 	\\
{Logic}	&	{0.849}		&	{2.810}	\\
{BRAM}	&	{0.436}		&	{1.308} 	\\
{PS7}	&	{1.623}		&	{1.625} 	\\
{Device Static}	&	{0.291}		&	{0.357} \\
\hline
\hline
{Total}	&	{4.982}		&	{9.830} \\
\noalign{\hrule height 2pt }
\end{widetable}
\end{table}

We compare this implementation with that of a high-end GPU, the NVIDIA GeForce Titan Black. The GPU can recognize 250K images at each second, which is about 3.6 times higher than that of this FPGA based implementation. Note that the GPU used floating-point arithmetic for this test.  But the accuracy difference is very minimal.  However, the GPU system not only demands a much bigger physical space but also consumes much higher power. The power consumption of the GPU system is about 250 Watt \cite{nvidiapower}.  Thus, the energy efficiency of this FPGA based system is considered to be over 10 times higher than that of the GPU based system. 

\begin{table}[htb]
\normalsize \centering
\renewcommand{\arraystretch}{1.6}
\caption{Hardware resources in Xilinx FPGA families.\newline \\ }
\label{tbl:dnn_kws_nkws}
\centering
\begin{widetable}{\columnwidth}{c | c | c | c | c }
\noalign {\vspace{1em}}
\noalign{\hrule height 2pt }
\thead{}		&	\thead{Kintex-7 }	&	\thead{Virtex-7}	&	\thead{Kintex\\ UltraScale}	&	\thead{Virtex\\ UltraScale}  \\
\hline
{LUTs (K)}	&	{299}	&	{1221} 	&	{663}	&	{2533}\\
{Block RAM (Mb)}	&	{34}	&	{68} 	  	&	{76}	&	{132.9}\\
{DSP slices}	&	{1920}	&	{3600}	&	{5520}	&	{2880}	\\
\noalign{\hrule height 2pt}
\end{widetable}
\end{table}

During the last a few years, many DNN algorithms have been developed.  Some of them is more complex than this design. A very complex DNN used for phoneme recognition demands 20 million weights \cite{mohamed2012acoustic}. Thus, this DNN circuit needs about 60 Mbits of BRAM to operate using 3-bit weights stored on on-chip memory.  The FPGA density is growing very rapidly as well.  Table 4 shows the hardware resources available in recent Xilinx FPGA families.  Note that the FPGA employed to this design is equivalent to the Kintex-7 family.  This table shows that even a very complex DNN algorithm can be implemented using an FPGA with only on-chip memory. More noticeable is the speed. With only 124K LUTs, the design achieved about 28\% speed of a high-end GPU based system. If we employ the highest density FPGA containing 2533K LUTs, it would be possible to acheive more than 5 times of the speed-up compared to the GPU based system. The IO bandwidth bound is not a limitation because no DRAM access is needed.

\section{Concluding Remarks}
\label{sec:conclusion}

We show an FPGA based implementation of a deep neural network for the handwritten digit recognition and the phoneme recognition problems, which needs millions of weights and arithmetic operations for producing one output.  In order to use only on-chip memory for weight storage, the weights are represented in 3 bits, while the internal signals employ the precision of 8 bits. A retrain based fixed-point weight optimization technique is employed to reduce the performance gap with floating-point algorithms.  The implemented FPGA shows the throughput that is about one quarter of a GPU based system, but only demands approximately 2\textasciitilde4\% of the power consumption of the GPU system, resulting in over 10 times of power efficiency. If we employ a larger FPGA, such as Virtex-7 or Virtex UltraScale, a much higher throughput than that of a GPU based system can be achieved.




%

\newpage

\bibliographystyle{IEEEbib}
\bibliography{IEEEabrv,icassp2016_jhpark}

\end{document}